\documentclass{article}
\usepackage{graphicx} 
\usepackage{fullpage}
\usepackage{amsmath}
\usepackage{amsfonts}
\usepackage{amsthm}
\usepackage{xcolor}
\usepackage{tikz}
\usepackage{subcaption}
\usepackage{pgfmath}
\usetikzlibrary{scopes}
\usetikzlibrary{backgrounds}
\usetikzlibrary{shapes}
\usetikzlibrary{calc}
\usetikzlibrary{automata}
\usetikzlibrary{decorations.pathmorphing}
\usetikzlibrary{shapes.symbols}
\usepackage{hyperref}

\newtheorem{theorem}{Theorem}

\newtheorem{lemma}{Lemma}

\theoremstyle{definition}

\usepackage{cleveref}

\usepackage{authblk}

\newcommand{\set}[1]{\left\{{#1}\right\}}

\DeclareMathOperator{\deficit}{\Delta}

\title{A Family of Well-Behaved Multiflow Instances\thanks{A version of this paper will appear in the proceedings of the 2027 SIAM Symposium on Simplicity in Algorithms (SOSA 2027).}}
\author[1]{Joseph Poremba}
\author[1]{F. Bruce Shepherd}
\affil[1]{Department of Computer Science, University of British Columbia, Canada}
\date{}

\begin{document}

\maketitle

\begin{abstract}
    We introduce a family of cut-sufficient instances for undirected multiflow that generalizes single-source planar and fully-planar instances. We also show the family has an additive $6 d_{\max}$ error for converting a fractional flow into an unsplittable one.
\end{abstract}


\section{Introduction}

Network flows are one of the most widely studied models of combinatorial optimization.
In the {\em multicommodity flow} problem, we are given a capacitated {\em supply graph} $G$, and asked to route flow between multiple designated source-sink pairs simultaneously while respecting the capacities.
The source-sink pairs are called {\em demands} or {\em commodities}, and are represented by a {\em demand graph} $H$.
There are two related models of multicommodity flow.
In the feasibility model, we are given specific values $(d(st): st \in E(H))$ of flow to route for each commodity.
In the maximization model, we are instead asked to maximize the total amount of flow routed, with no specific requirements on the values for each commodity.
This paper primarily focuses on the feasibility model.
All graphs in this paper are undirected.

The single-source version of multicommodity flow (i.e. all commodities share a single common source) famously exhibits many nice combinatorial properties.
\begin{enumerate}
    \item Max-flow min-cut property (also called {\em cut-sufficiency}): if every cut has at least as much capacity crossing it as demand crossing it (this is the so-called {\em cut condition}), then the instance is feasible \cite{ford_maximal_1956}.
    \item Integrality: provided the capacity and demand values are integral, if an instance is feasible then it has an integral feasible flow \cite{ford_maximal_1956}.
    \item Low-error unsplittable flow: any feasible instance has a flow that is {\em unsplittable} (routes each commodity on only one path) and violates the edge capacities by an additive error of only $d_{\max} = \max \set{d(st) : st \in E(H)}$ \cite{Dinitz99}.
\end{enumerate}

General multiflows do not exhibit these properties.
In fact, they can be violated quite badly.
Classical work has introduced several different cut-sufficient families (see \cite{schrijver2003combinatorial} for a catalogue) and each of these has come with a half-integrality guarantee. One famous example is the family of fully-planar instances \cite{seymour1981odd}, where $G+H$ is planar.
In this work, we present a new family of cut-sufficient multiflow instances that generalizes both planar single-source and fully-planar instances.
We call these {\em stellar-planar} instances and we also provide a half-integrality result for integer data inputs.

Recent advances have also been made in low-error unsplittable flow for several cut-sufficient families, in particular $2 d_{\max}$ error for outerplanar instances \cite{aleman_espinosa_pinning_2026} (improved from $\frac{18}{5} d_{\max}$ \cite{aleman2025unsplittable}) and $2 d_{\max}$ error for fully-planar instances \cite{espinosa2026unsplittable}.
We establish a $6d_{\max}$ additive error for unsplittable flow in stellar-planar instances.

Our techniques are inspired by previous work on cut-sufficiency in fully-compliant series-parallel instances \cite{chekuri2013flow}, along with unsplittable flow in outerplanar networks \cite{aleman2025unsplittable} and series-parallel and fully-planar instances \cite{espinosa2026unsplittable}, despite these classes not immediately resembling stellar-planar instances.
However, due to the nice structure of stellar-planar instances, our proofs for these applications are ultimately much shorter and we believe easier to understand, and therefore provide a good entry point for a reader to understand these lines of work in more depth.
At a very high level, these proofs all somehow involve splitting an instance at a 2-cut and compressing one or both sides to simplify the instance.

\subsection{Related Work}
\label{sec:related}

There has been intensive research in establishing families of undirected multiflow instances for which the cut condition is sufficient to guarantee feasibility. The earliest successes were for instances with one or two demands \cite{ford_maximal_1956,hu_multi-commodity_1963}. The former also inspired work on single-source demand instances. 
Instances with at most four terminals are also cut-sufficient \cite{seymour1980four,lomonosov1985combinatorial}. In the 1980s topological cut-sufficient instances were introduced, such as fully-planar \cite{seymour1981odd} and planar instances where demands span one or two faces  \cite{Okamura81} and \cite{okamura1983multicommodity}.
More recently cut-sufficiency in series-parallel graphs was characterized in \cite{chakrabarti2012cut}, settling a conjecture in \cite{chekuri2013flow}. 




\section{Stellar-Planar Instances and Results}

To motivate our new class of multiflow instances, consider the following equivalent definition of fully-planar instances.
Let $K$ be a planar {\em base graph} (possibly with parallel edges) defined on node set $V_0$.
A fully-planar multiflow instance is obtained from $K$ by replacing every edge $e = a b$ with either a supply edge or a demand edge joining $a,b$, or both.

\begin{figure}[tb]
    \centering
    \begin{subfigure}[b]{0.45\textwidth}
        \centering
        
\begin{tikzpicture}[scale=1.0,
    vertex/.style={circle, fill=black, inner sep=3pt}
]
\node[vertex] (v1) at (-2.5,0) {};
\node[vertex] (v2) at (0,5) {};
\node[vertex] (v3) at (-5,5) {};
\draw (v1) -- (v2) -- (v3) -- (v1);

\end{tikzpicture}
        \caption{The base graph $K$.}
        \label{fig:sub1}
    \end{subfigure}
    \hfill
    \begin{subfigure}[b]{0.45\textwidth}
        \centering
\begin{tikzpicture}[scale=1.0,
    vertex/.style={circle, fill=black, inner sep=3pt}
]

\node[vertex] (v1) at (-2.5,0) {};
\node[vertex] (v2) at (0,5) {};
\node[vertex] (v3) at (-5,5) {};

\node[cloud,draw,cloud puffs=8, cloud puff arc=60,minimum width = 5.5 cm, minimum height = 1 cm,rotate=65,fill=lightgray] (c1) at ($(v1)!0.5!(v2)$) { };
\node[cloud,draw,cloud puffs=8, cloud puff arc=60,minimum width = 5 cm, minimum height = 1 cm,rotate=0,fill=lightgray] (c2) at ($(v2)!0.5!(v3)$) { };
\node[cloud,draw,cloud puffs=8, cloud puff arc=60,minimum width = 5.5 cm, minimum height = 1 cm,rotate=-65,fill=lightgray] (c3) at ($(v3)!0.5!(v1)$) { };

\node[vertex] (v1') at (-2.5,0) {};
\node[vertex] (v2') at (0,5) {};
\node[vertex] (v3') at (-5,5) {};

\node (t12a) at ($(v1)!0.33!(v2)$) { };
\node (t12b) at ($(v1)!0.67!(v2)$) { };

\node (t23a) at ($(v2)!0.33!(v3)$) { };
\node (t23b) at ($(v2)!0.67!(v3)$) { };

\node (t31a) at ($(v3)!0.33!(v1)$) { };
\node (t31b) at ($(v3)!0.67!(v1)$) { };

\draw[red,dashed,thick] (v1) to[bend right=60] (t12a);
\draw[red,dashed,thick] (v1) to[bend right=60] (t12b);
\draw[red,dashed,thick] (v2) to[bend right=60] (t23a);
\draw[red,dashed,thick] (v2) to[bend right=60] (t23b);
\draw[red,dashed,thick] (v3) to[bend right=60] (t31a);
\draw[red,dashed,thick] (v3) to[bend right=60] (t31b);


\end{tikzpicture}
        \caption{The stellar-planar multiflow instance.}
        \label{fig:sub2}
    \end{subfigure}
    \caption{A schematic showing how to create a stellar-planar multiflow instance from a planar base graph. Each gray cloud is some planar supply graph. The red dashed edges are demand edges.}
    \label{fig:stellar-planar}
\end{figure}
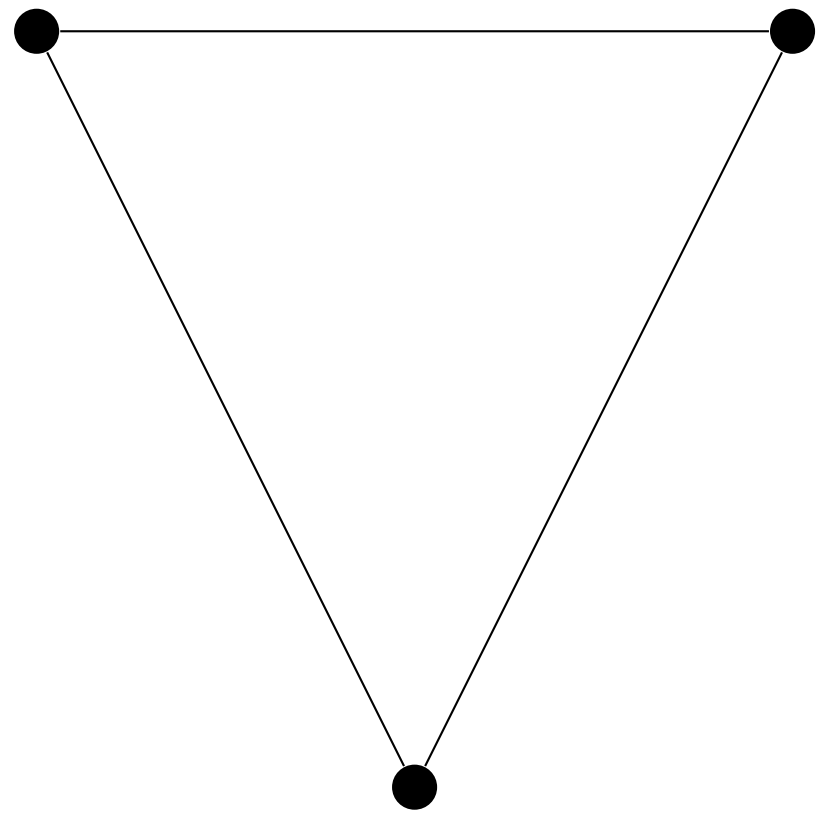
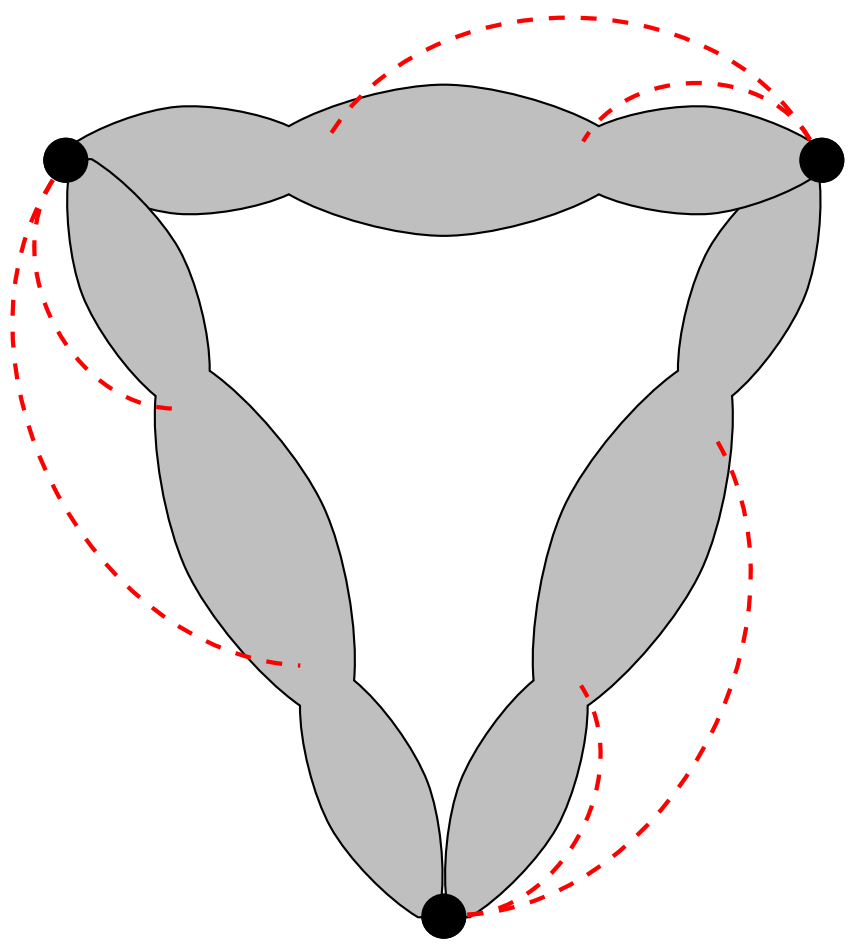

\begin{figure}[tb]
\centering
\definecolor{spe1}{rgb}{0.74,0.84,1.00}
\definecolor{spe2}{rgb}{0.76,0.92,0.72}
\definecolor{spe3}{rgb}{0.98,0.68,0.45}
\definecolor{spe4}{rgb}{0.86,0.77,0.96}
\definecolor{spe5}{rgb}{0.68,0.91,0.90}
\definecolor{spe6}{rgb}{1.00,0.96,0.45}
\tikzset{
  att/.style={circle, fill=black, inner sep=0pt, minimum size=6pt},
  iv/.style={circle, fill=black, inner sep=0pt, minimum size=3.6pt},
  supply/.style={black, line width=0.7pt},
  demand/.style={red, line width=1.2pt, dash pattern=on 3.5pt off 2pt},
  lbl/.style={font=\small},
  hlwidth/.initial=5pt,
  hl/.style={draw=sp#1, line width=\pgfkeysvalueof{/tikz/hlwidth}, line cap=round, line join=round, solid},
}
\begin{subfigure}[b]{0.30\textwidth}
\centering
\begin{tikzpicture}[scale=0.78]
\coordinate (a) at (0,0);
\coordinate (c) at (5,0);
\coordinate (b) at (2.5,4);
\coordinate (z) at (2.5,1.5);
\begin{scope}[hlwidth=7pt]
\draw[hl=e1] (a) -- (b);
\draw[hl=e2] (b) -- (c);
\draw[hl=e3] (a) -- (c);
\draw[hl=e4] (a) -- (z);
\draw[hl=e5] (z) -- (c);
\draw[hl=e6] (z) -- (b);
\end{scope}
\draw[supply] (a) -- node[left=3pt, font=\scriptsize] {$e_1$} (b)
              -- node[right=3pt, font=\scriptsize] {$e_2$} (c)
              -- node[below=3pt, font=\scriptsize] {$e_3$} (a);
\draw[supply] (a) -- node[above=3pt, pos=0.45, font=\scriptsize] {$e_4$} (z)
              -- node[above=3pt, pos=0.55, font=\scriptsize] {$e_5$} (c);
\draw[supply] (z) -- node[right=2pt, font=\scriptsize] {$e_6$} (b);
\foreach \v/\p in {a/below left, b/above, c/below right, z/below}
  \node[att, label={[lbl, inner sep=2pt]\p:$\v$}] at (\v) {};
\end{tikzpicture}
\caption{The base graph $K$.}
\label{fig:stellar-base}
\end{subfigure}%
\hfill
\begin{subfigure}[b]{0.66\textwidth}
\centering
\begin{tikzpicture}[scale=0.68]
\node[att, label={[lbl]below left:$a$}] (a) at (0,0) {};
\node[att, label={[lbl]below right:$c$}] (c) at (10,0) {};
\node[att, label={[lbl]above:$b$}] (b) at (5,10) {};
\node[att, label={[lbl, inner sep=2.5pt]128:$z$}] (z) at (5,3.4) {};
\node[iv] (e1L) at (3.93,6.39) {};
\node[iv] (e1X) at (0.43,3.78) {};
\node[iv] (e1M) at (2.64,7.96) {};
\node[iv] (e1R) at (1.97,9.75) {};
\node[iv] (e1Y) at (-1.07,4.26) {};
\node[iv] (e1W) at (-2.57,4.75) {};
\node[iv] (e2S) at (9.11,5.80) {};
\node[iv] (e2P) at (10.05,7.94) {};
\node[iv] (e2Q) at (13.32,4.99) {};
\node[iv] (e2W) at (10.46,5.57) {};
\node[iv] (e2T) at (11.89,2.71) {};
\node[iv] (e4U) at (1.73,2.07) {};
\node[iv] (e4V) at (2.99,2.19) {};
\node[iv] (e4m) at (1.86,1.36) {};
\node[iv] (e4n) at (2.69,0.78) {};
\node[iv] (e4p) at (3.62,3.56) {};
\node[iv] (e4q) at (4.15,3.23) {};
\node[iv] (e4r) at (3.82,2.78) {};
\node[iv] (e5O1) at (7.80,3.25) {};
\node[iv] (e5O2) at (6.45,1.26) {};
\node[iv] (e5I1) at (6.31,2.75) {};
\node[iv] (e5I2) at (7.49,2.61) {};
\node[iv] (e5I3) at (8.06,1.50) {};
\node[iv] (e5I4) at (6.84,1.66) {};
\begin{scope}[on background layer]
  \draw[hl=e1] (a) -- (e1L);
  \draw[hl=e1] (e1L) -- (b);
  \draw[hl=e1] (b) -- (e1R);
  \draw[hl=e1] (e1R) -- (e1W);
  \draw[hl=e1] (e1W) -- (a);
  \draw[hl=e1] (b) .. controls (2.59,11.88) and (-1.97,9.26) .. (e1W);
  \draw[hl=e1] (e1L) -- (e1M);
  \draw[hl=e1] (e1M) -- (e1R);
  \draw[hl=e1] (e1M) -- (e1X);
  \draw[hl=e1] (a) -- (e1X);
  \draw[hl=e1] (e1X) -- (e1R);
  \draw[hl=e1] (e1L) -- (e1X);
  \draw[hl=e1] (e1X) -- (e1Y);
  \draw[hl=e1] (e1Y) -- (e1W);
  \draw[hl=e2] (b) -- (e2P);
  \draw[hl=e2] (e2P) -- (e2Q);
  \draw[hl=e2] (b) -- (e2S);
  \draw[hl=e2] (e2S) -- (c);
  \draw[hl=e2] (e2P) -- (e2W);
  \draw[hl=e2] (e2W) -- (e2S);
  \draw[hl=e2] (e2W) -- (e2T);
  \draw[hl=e2] (e2Q) -- (e2T);
  \draw[hl=e2] (e2T) -- (c);
  \draw[hl=e3] (a) -- (c);
  \draw[hl=e4] (a) -- (e4U);
  \draw[hl=e4] (e4U) -- (e4V);
  \draw[hl=e4] (e4U) -- (e4m);
  \draw[hl=e4] (a) -- (e4m);
  \draw[hl=e4] (e4m) -- (e4V);
  \draw[hl=e4] (e4m) -- (e4n);
  \draw[hl=e4] (a) -- (e4n);
  \draw[hl=e4] (e4V) -- (e4n);
  \draw[hl=e4] (z) -- (e4p);
  \draw[hl=e4] (e4p) -- (e4r);
  \draw[hl=e4] (e4p) -- (e4q);
  \draw[hl=e4] (z) -- (e4q);
  \draw[hl=e4] (e4q) -- (e4r);
  \draw[hl=e5] (z) -- (e5O1);
  \draw[hl=e5] (e5O1) -- (c);
  \draw[hl=e5] (c) -- (e5O2);
  \draw[hl=e5] (e5O2) -- (z);
  \draw[hl=e5] (e5I1) -- (e5I2);
  \draw[hl=e5] (e5I2) -- (e5I3);
  \draw[hl=e5] (e5I3) -- (e5I4);
  \draw[hl=e5] (e5I4) -- (e5I1);
  \draw[hl=e5] (z) -- (e5I1);
  \draw[hl=e5] (e5O1) -- (e5I2);
  \draw[hl=e5] (c) -- (e5I3);
  \draw[hl=e5] (e5O2) -- (e5I4);
  \draw[hl=e5] (z) .. controls (5.30,0.29) and (7.63,0.34) .. (c);
  \draw[hl=e1] (a) -- (b);
  \draw[hl=e1] (a) -- (e1M);
  \draw[hl=e1] (a) .. controls (-0.54,2.84) and (-0.40,6.24) .. (e1R);
  \draw[hl=e2] (b) -- (e2W);
  \draw[hl=e2] (b) .. controls (10.99,10.54) and (12.74,6.60) .. (e2T);
  \draw[hl=e4] (z) -- (e4r);
  \draw[hl=e4] (z) -- (e4n);
  \draw[hl=e4] (z) .. controls (4.05,2.50) and (3.05,1.55) .. (e4m);
  \draw[hl=e5] (c) .. controls (8.94,2.80) and (7.93,5.76) .. (z);
  \draw[hl=e5] (c) -- (e5I2);
  \draw[hl=e5] (c) -- (e5I4);
  \draw[hl=e6] (b) -- (z);
\end{scope}
\draw[supply] (a) -- (e1L);
\draw[supply] (e1L) -- (b);
\draw[supply] (b) -- (e1R);
\draw[supply] (e1R) -- (e1W);
\draw[supply] (e1W) -- (a);
\draw[supply] (b) .. controls (2.59,11.88) and (-1.97,9.26) .. (e1W);
\draw[supply] (e1L) -- (e1M);
\draw[supply] (e1M) -- (e1R);
\draw[supply] (e1M) -- (e1X);
\draw[supply] (a) -- (e1X);
\draw[supply] (e1X) -- (e1R);
\draw[supply] (e1L) -- (e1X);
\draw[supply] (e1X) -- (e1Y);
\draw[supply] (e1Y) -- (e1W);
\draw[supply] (b) -- (e2P);
\draw[supply] (e2P) -- (e2Q);
\draw[supply] (b) -- (e2S);
\draw[supply] (e2S) -- (c);
\draw[supply] (e2P) -- (e2W);
\draw[supply] (e2W) -- (e2S);
\draw[supply] (e2W) -- (e2T);
\draw[supply] (e2Q) -- (e2T);
\draw[supply] (e2T) -- (c);
\draw[supply] (a) -- (c);
\draw[supply] (a) -- (e4U);
\draw[supply] (e4U) -- (e4V);
\draw[supply] (e4U) -- (e4m);
\draw[supply] (a) -- (e4m);
\draw[supply] (e4m) -- (e4V);
\draw[supply] (e4m) -- (e4n);
\draw[supply] (a) -- (e4n);
\draw[supply] (e4V) -- (e4n);
\draw[supply] (z) -- (e4p);
\draw[supply] (e4p) -- (e4r);
\draw[supply] (e4p) -- (e4q);
\draw[supply] (z) -- (e4q);
\draw[supply] (e4q) -- (e4r);
\draw[supply] (z) -- (e5O1);
\draw[supply] (e5O1) -- (c);
\draw[supply] (c) -- (e5O2);
\draw[supply] (e5O2) -- (z);
\draw[supply] (e5I1) -- (e5I2);
\draw[supply] (e5I2) -- (e5I3);
\draw[supply] (e5I3) -- (e5I4);
\draw[supply] (e5I4) -- (e5I1);
\draw[supply] (z) -- (e5I1);
\draw[supply] (e5O1) -- (e5I2);
\draw[supply] (c) -- (e5I3);
\draw[supply] (e5O2) -- (e5I4);
\draw[supply] (z) .. controls (5.30,0.29) and (7.63,0.34) .. (c);
\draw[demand] (a) -- (b);
\draw[demand] (a) -- (e1M);
\draw[demand] (a) .. controls (-0.54,2.84) and (-0.40,6.24) .. (e1R);
\draw[demand] (b) -- (e2W);
\draw[demand] (b) .. controls (10.99,10.54) and (12.74,6.60) .. (e2T);
\draw[demand] (z) -- (e4r);
\draw[demand] (z) -- (e4n);
\draw[demand] (z) .. controls (4.05,2.50) and (3.05,1.55) .. (e4m);
\draw[demand] (c) .. controls (8.94,2.80) and (7.93,5.76) .. (z);
\draw[demand] (c) -- (e5I2);
\draw[demand] (c) -- (e5I4);
\draw[demand] (b) -- (z);
\end{tikzpicture}
\caption{The stellar-planar multiflow instance.}
\label{fig:stellar-instance}
\end{subfigure}
\caption{An example showing a stellar-planar multiflow instance.
The background colour of each patch matches the
edge of $K$ it replaces.
In the multiflow instance, black edges are supply edges and red dashed edges are demand edges.
The patch corresponding to $e_3$ is a single supply edge (trivial) while the patch corresponding to $e_6$ is a single demand edge (trivial).
The supply graph of the
patch corresponding to $e_4$ is disconnected.}
\label{fig:stellar-planar-concrete}
\end{figure}
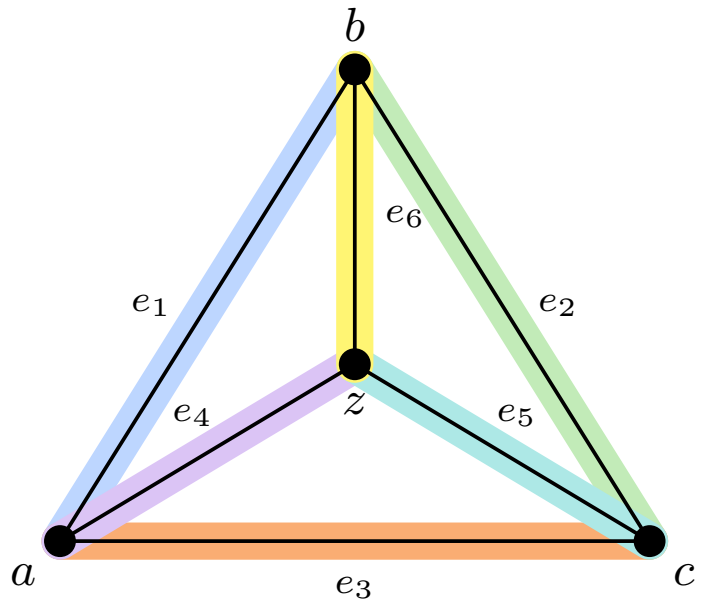
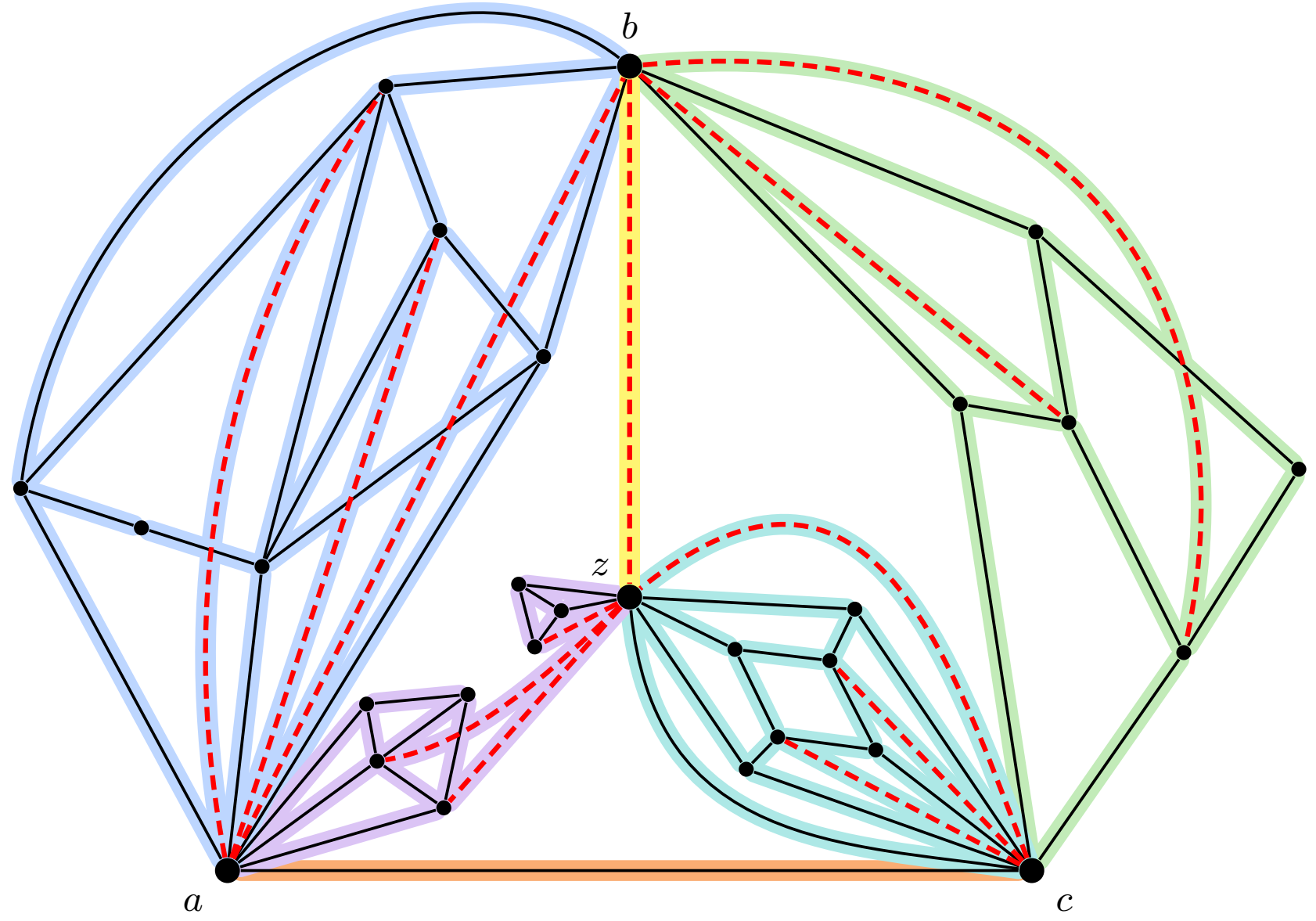


Stellar-planar instances generalize this by allowing $e = ab$ to be replaced by an arbitrary planar single-source instance with $a$ as its source and both $a$ and $b$ on its outside face\footnote{``Stellar'' refers to the fact that each sub-instance has a star demand graph, and so the whole instance is a `constellation' of single-source instances.}.
\Cref{fig:stellar-planar} shows a basic schematic example of how to construct such an instance, and \Cref{fig:stellar-planar-concrete} shows a more complicated, concrete instance.
We call each of these sub-instances corresponding to an edge of $K$ a {\em patch}.
We require that each patch is node-disjoint from the others except for its associated {\em attachment nodes} $\set{a, b}$ and that the endpoints of all demands associated with a patch must lie within that patch.
Note that the supply graph of a single patch does not need to be connected
and that the whole supply graph is the edge-disjoint union of these patch supply graphs.
The demand graph similarly is the edge-disjoint union of the patch demand graphs.

A patch is {\em trivial} if its only nodes are $a$ and $b$.
A fully-planar instance is a stellar-planar instance constructed with just trivial patches, while a planar single-source instance is a stellar-planar instance constructed from a single patch\footnote{Provided that its source node is on the outside face, which can be assumed without loss of generality.}.

\subsection{Results}
Before we state our results, we establish more precise definitions.
A {\em multiflow instance} consists of $(G, H, u, d)$, where $G$ is a supply graph defined on node set $V$, $H$ is a demand graph on $V$, $u$ are edge capacities on $G$, and $d$ are demand values on edges of $H$.
We allow multiple edges in $G$ and $H$, but frequently just use $st$ to denote an edge (supply or demand) with endpoints $s,t$.
For a set $R$ of edges, we let $u(R) = \sum_{e \in R} u(e)$ and use similar notation for summing $d$ over sets.
For $h = st \in E(H)$, let $\mathcal P_{h}$ denote the set of $st$-paths in $G$.
An $h$-flow is a function $x^h: \mathcal P_{h} \to \mathbb R_{\ge 0}$.
A multiflow is a collection $(x^h : h \in E(H))$.
Multiflows can be fractional unless otherwise stated.
We let $x(e) = \sum_{h \in E(H)} \sum_{P \in \mathcal P_h : e \in P} x^h(P)$ which is the sum of flow on edge $e \in E(G)$, and let $|x^{h}| = \sum_{P \in \mathcal P_{h}} x^h(P)$ which is the amount of demand $h$ that is routed.
A multiflow is feasible if $x(e) \le u(e)$ for all $e \in E(G)$ and $|x^{h}| \ge d(h)$ for all $h \in E(H)$.

For simplicity we state and prove our results as existential theorems, but it is straightforward to convert each into a polynomial time algorithm.
Our first result concerns cut-sufficiency and half-integrality.
We say an instance $(G, H, u, d)$ satisfies the {\em cut condition} if $u(\delta_G(X)) \ge d(\delta_H(X))$ for all $X \subseteq V$, where $\delta(X)$ refers to the set of edges with one endpoint in $X$ and one endpoint in $V \setminus X$.
The cut condition is necessary, but not in general sufficient for an instance to be feasible.
A multiflow $x$ is {\em half-integral} if $2x$ is integral.
We say the instance satisfies the {\em Euler condition} if for every node $v \in V$, $\sum_{e \in \delta_G(v)} u(e) + \sum_{h \in \delta_H(v)}d(h)$ is an even integer.
\begin{theorem}[Section~\ref{sec:cs}]
    \label{thm:cut-sufficient}
    Let $(G, H, u, d)$ be a stellar-planar multiflow instance.
    If the cut condition is satisfied, then the instance is feasible.
    If additionally $u, d$ are integral, then there exists a feasible half-integral flow.
    If also the Euler condition is satisfied, then there exists a feasible integral flow.
\end{theorem}

The next result concerns unsplittable flow.
A flow $x$ is {\em unsplittable} if it satisfies all demands and, for each $h \in E(H)$, exactly one path $P \in \mathcal P_{h}$ has $x^h(P) > 0$.
We let $d_{\max} = \max_{h \in E(H)} d(h)$.
\begin{theorem}[Section~\ref{sec:usf}]
    \label{thm:unsplittable}
    Let $(G, H, u, d)$ be a stellar-planar multiflow instance.
    If the instance is feasible, then there exists an unsplittable flow $y$ that routes all demands such that $y(e) \le u(e) + 6d_{\max}$ for all $e \in E(G)$.
\end{theorem}



\subsection{Notation for Patches}
Consider a stellar-planar multiflow instance $(G, H, u, d)$ obtained from a base graph $K$.
For a patch $C$ replacing edge $ab$ of $K$, we identify $C$ with a tuple $(B, F, a, b)$, where $B$ is the supply graph, $F$ is the demand graph, and we assume all edges of $F$ are incident with $a$.
We let $\overline B$ be the supply graph obtained from $G$ by deleting $E(B)$ and all nodes of $B$ except for $a, b$.
We let $\overline F = E(H) - F$.




\section{Cut-Sufficiency}
\label{sec:cs}

To prove \Cref{thm:cut-sufficient}, it suffices to prove the last statement where we assume the Euler condition.
For half-integrality, if the Euler condition is not satisfied we can double both $u$ and $d$, then scale down the resulting integer flow. Similarly for fractional flows we can scale (rational) demands and capacities to become integral and Eulerian.

For a multiflow instance $\mathcal I = (G, H, u, d)$, we define the {\em deficit} of a set $S \subseteq V$ to be $\deficit(\mathcal I, S) = d(\delta_H(S)) - u(\delta_G(S))$, which is the amount by which the cut condition is violated for this cut.
The {\em deficit} of the entire instance $\mathcal I$ is $\deficit(\mathcal I) = \max_{S \subseteq V} \deficit(\mathcal I, S)$.
The cut condition is equivalent to $\deficit(\mathcal I) \le 0$.
For two nodes $a, b$, the {\em $ab$-deficit} is defined as $\deficit_{ab}(\mathcal I) = \max_{S \subseteq V : a \in S, b \notin S} \deficit(\mathcal I, S)$.

We use the high-level strategy of {\em 2-cut reduction}, used by \cite{chekuri2013flow} for fully-compliant series-parallel instances.
We explain how it works in our setting, before presenting the general lemma.
For us, the 2-cut in question is $\set{a, b}$ for a patch $C = (B, F, a, b)$.
Let $\mathcal I(C)$ denote the multiflow instance $(B, F, u_C, d_C)$ where $u_C$ is $u$ restricted to $E(B)$ and $d_C$ is $d$ restricted to $F$.
Let $\mathcal I(\overline C)$ denote the multiflow instance $(\overline B, \overline F, u_{\overline C}, d_{\overline C})$, where $u_{\overline C}, d_{\overline C}$ are defined analogously.
For notational convenience we let $\deficit_{ab}(C) = \deficit_{ab}(\mathcal I(C))$ and $\deficit_{ab}(\overline C) = \deficit_{ab}(\mathcal I(\overline C))$.

Even if the overall instance $\mathcal I = (G, H, u, d)$ satisfies the cut condition, it is still possible that $\mathcal I(C)$ or $\mathcal I(\overline C)$ does not, so we cannot just immediately apply induction.
Any cut condition violation for one of the smaller instances arises from some $S \subseteq V$ where $a \in S, b \notin S$.
However, if say $\mathcal I(C)$ has a positive $ab$-deficit $k$, then it is easy to show that $\mathcal I(\overline C)$ cannot have positive $ab$-deficit (else the cut condition would fail for $\mathcal I$).
In fact, $\mathcal I(\overline C)$ must have excess capacity to compensate for $\mathcal I(C)$'s $ab$-deficit.
We can take advantage of this to take the demands that $\mathcal I(C)$ cannot route within $B$ and route them through $\overline B$ instead.
We simulate this spare capacity that $\mathcal I(C)$ can use in $\mathcal I(\overline C)$ by adding an extra supply edge $ab$ to $\mathcal I(C)$ and obtain a new instance $\mathcal L(C)$.
Similarly, we simulate the extra demand that $\mathcal I(\overline C)$ has to accommodate by adding an extra demand edge $ab$ and obtaining a new instance $\mathcal L(\overline C)$.
We then inductively find integral flows for $\mathcal L(C), \mathcal L(\overline C)$ and show they can be glued together into a flow for the original instance.

Now we present the generic 2-cut reduction lemma from \cite{chekuri2013flow}.
For any distinct nodes $a,b$ (not necessarily attachment nodes of a stellar-planar instance), an $\set{a,b}$-partition of $G$ is any pair of graphs $(G_1, G_2)$ such that: (1) $V(G_1) \cap V (G_2) = \set{a, b}$, and (2) $E(G)$ is the disjoint union of $E(G_1), E(G_2)$.

\begin{lemma}[2-Cut Reduction, Lemma 3.3 of \cite{chekuri2013flow}]
    \label{lemma:2cut}
    Let $(G, H, u, d)$ satisfy the cut condition where $u, d$ are integral.
    Let $a, b$ be distinct nodes in $G$.
    Let $G_1, G_2$ be an $\set{a, b}$-partition.
    Suppose $H$ has no edges between $G_1$ and $G_2$ --- that is, its edges can be partitioned into $H_1, H_2$ such that each edge of $H_i$ has both endpoints in $G_i$.
    Let $u_i, d_i$ be the restriction of $u, d$ to $G_i, H_i$.
    It is straightforward to check that the $ab$-deficit of at least one of the reduced instances $(G_i, H_i, u_i, d_i)$ is $\le 0$.
    Let $k_i$ be the $ab$-deficit of $(G_i, H_i, u_i, d_i)$.
    Suppose at least one $k_i$ is nonnegative\footnote{In \cite{chekuri2013flow} the deficit is defined to always be $\ge 0$ by (essentially) taking the $\max$ of our deficit definition and 0. However, there is a small bug in their Lemma 3.3 in the event both instances have strictly negative $ab$-deficit (or in their definition, would be negative but take value 0 from the $\max$). There is a parity counting issue that causes the Euler condition to not be preserved. We thank an anonymous reviewer for pointing this out. The version stated here is corrected since we assume $k_1$ is nonnegative. The case of two negative deficits is easy to handle separately, in their work and ours --- we handle it directly in \Cref{thm:cut-sufficient-integral-euler}.}; without loss of generality $k_1 \ge 0 \ge k_2$.
    Let $H'_2, d'_2$ be obtained by adding $k_1$ units of demand on edge $ab$ in $H_2$ (adding this edge if it does not exist).
    Let $G'_1, u'_1$ be obtained by adding $k_1$ units of capacity on edge $ab$ to $G_1$ (adding this edge if it does not exist).
    We let $H'_1 = H_1, d'_1 = d_1, G'_2 = G_2, u'_2 = u_2$.
    Then $(G'_i, H'_i, u'_i, d'_i)$ satisfies the cut condition for $i = 1, 2$.
    Moreover, if $(G, H, u, d)$ satisfies the Euler condition, then so too does each $(G'_i, H'_i, u'_i, d'_i)$.
    Finally, if there is a feasible integral flow for each instance $(G'_i, H'_i, u'_i, d'_i)$, then there is a feasible integral flow for $(G, H, u, d)$.
\end{lemma}

In our case, $(G_1, H_1, u_1, d_1), (G_2, H_2, u_2, d_2)$ are $\mathcal I(C)$ and $\mathcal I(\overline C)$ (in some order).
There is a catch to this strategy --- how do we inductively find integral flows for $(G'_1, H'_1, u'_1, d'_1)$, $(G'_2, H'_2, u'_2, d'_2)$?
It may not be the case that these instances are smaller than $\mathcal I$, for example if $G_2$ consists of a single edge. So induction may not be immediately applicable.
In \cite{chekuri2013flow}, they argue that a 2-cut exists in a fully-compliant instance such that $G_1, G_2$ are not just single edges. 
We do something different. If we cannot make progress, it is because our patches are all trivial.
Therefore we have a fully-planar instance, and we are already done!
We now formalize this proof as follows.

\begin{theorem}
    \label{thm:cut-sufficient-integral-euler}
    Let $(G, H, u, d)$ be a stellar-planar instance that satisfies the cut condition, where $u, d$ are integral and satisfy the Euler condition.
    There exists a feasible integral multiflow.
\end{theorem}
\begin{proof}
    Suppose the theorem is false, and consider a counterexample $\mathcal I = (G, H, u, d)$ minimizing the number of non-trivial patches.
    Observe that if every patch is trivial, then the instance is fully-planar and thus the result follows by Seymour's work \cite{seymour1981odd}.
    So suppose there is a non-trivial patch $C = (B, F, a, b)$.

    First, note that we may assume $\deficit_{ab}(\mathcal I) = 0$.
    If this is not the case, then we may add a demand edge $ab$ with demand value $|\deficit_{ab}(\mathcal I)|$ without affecting the cut condition or the Euler condition (note that $\deficit_{ab}(\mathcal I)$ is even since $\mathcal I$ has the Euler condition).
    This does not increase the number of non-trivial patches.
    
    Now, as stated in \Cref{lemma:2cut}, the $ab$-deficit of at least one of the instances $\mathcal I(C), \mathcal I(\overline C)$ is $\le 0$.
    However, it cannot be the case that both are strictly negative, since then $\deficit_{ab}(\mathcal I) < 0$, contradicting the above assumption.
    We split into cases.

    Case 1: $\deficit_{ab}(C) \le 0$ and $\deficit_{ab}(\overline C) \ge 0$. Let $k = \deficit_{ab}(\overline C)$.
    Consider the following two instances.
    \begin{itemize}
        \item Let $\mathcal L(C)$ be obtained from $\mathcal I(C)$ by adding a demand edge $ab$ with demand value $k$ (or increasing the demand value by $k$ if it already exists).
        \item Let $\mathcal L(\overline C)$ be obtained from $\mathcal I(\overline C)$ by adding a supply edge $ab$ with capacity $k$ (or increasing the capacity by $k$ if it already exists).
    \end{itemize}
    By \Cref{lemma:2cut}, both of these instances satisfy the cut condition and Euler condition, and it suffices to find an integral flow for each of them.
    Moreover, they are still stellar-planar --- notice that $\mathcal L(\overline C)$ is obtained from the original instance by replacing $C$ with a trivial patch consisting of a single supply edge and zero demand edges.
    Since $\mathcal L(C)$ is single-source, it also has a feasible integral flow because the cut condition holds.
    Additionally, $\mathcal L(\overline C)$ has one fewer non-trivial patch than the original instance $\mathcal I$, so by minimality it has a feasible integral flow.
    The result follows.

    Case 2: $\deficit_{ab}(\overline C) \le 0$ and $\deficit_{ab}(C) \ge 0$. Let $k = \deficit_{ab}(C)$.
    Consider the following two instances.
    \begin{itemize}
        \item Let $\mathcal L(C)$ be obtained from $\mathcal I(C)$ by adding a supply edge $ab$ with capacity $k$ (or increasing the capacity by $k$ if it already exists).
        \item Let $\mathcal L(\overline C)$ be obtained from $\mathcal I(\overline C)$ by adding a demand edge $ab$ with value $k$ (or increasing the demand value by $k$ if it already exists).
    \end{itemize}
    As before, it suffices to find an integral flow for these two instances by \Cref{lemma:2cut}.
    The former is again a single-source instance and this is easy.
    For the latter, it is stellar-planar with one fewer non-trivial patch than $\mathcal I$ because it has replaced $C$ with a trivial patch consisting of a single demand edge and zero supply edges.
    Thus it has a feasible integral flow by minimality and the result follows.
\end{proof}

The only challenge in converting the above into a polynomial time algorithm is computing the $ab$-deficits for $\mathcal I(C)$ and $\mathcal I(\overline C)$.
In either case, first note that we can check the cut condition in polynomial time, since by the (existential) \Cref{thm:cut-sufficient}, the cut condition is equivalent to the existence of a feasible multiflow (which can be checked by linear programming).
This is true even if we were to add new supply or demand edges between $a$ and $b$.
If the cut condition holds, we have that $|\deficit_{ab}|$ is the maximum amount of demand we can put on a new edge $ab$ while still satisfying the cut condition.
A binary search can then find the value.
If the cut condition fails (only for $\mathcal I(C), \mathcal I(\overline C)$ since $\mathcal I$ satisfies the cut condition), then we know $\deficit_{ab} > 0$ and $\deficit_{ab}$ is the minimum amount of capacity we must put on this edge in order to satisfy the cut condition --- again we can employ a binary search.

\section{Unsplittable Flows}
\label{sec:usf}

To prove \Cref{thm:unsplittable}, we use an operation called pinning.
To pin a demand $h = st$ is to replace it with a sequence of demands $s y_1, y_1y_2, \dotsc, y_k t$ and assign each a demand value equal to $d(st)$.
Any $(H', d')$ obtained from $(H, d)$ using such operations is called a {\em pinning} of $(H, d)$.
Of course, any multiflow feasible for $(G, H', u, d')$ is also feasible for $(G, H, u, d)$ (after short-circuiting into simple paths, if necessary).
Our goal is to pin our instance into one of a nicer form (for which we already know how to compute low-congestion unsplittable flows), perhaps at the cost of increasing the capacities slightly to maintain feasibility.
A pinning strategy was also used before in \cite{aleman2025unsplittable,espinosa2026unsplittable,aleman_espinosa_pinning_2026}.

In our case, we find a pinning such that $G+H'$ is planar and the resulting instance is still feasible if we increase capacities by $4d_{\max}$.
We then invoke a result of \cite{espinosa2026unsplittable} which says that any fully-planar instance that is feasible has an unsplittable flow with additive congestion $2d_{\max}$.
Combining these, we obtain a total additive increase of $6d_{\max}$.


Let $x$ be a feasible flow for the instance $(G, H, u, d)$.
Consider a patch $C = (B, F, a, b)$.
We let $x_C$ denote the flow routing demands in patch $C$ and $x_{\overline C}$ denote the flow routing the demands from all other patches (so $x$ is the sum of $x_C$ and $x_{\overline C}$).
A support path $P$ of $x_C$ is {\em local} if $P$ is contained in $E(B)$.
Otherwise, it is {\em external}, and it consists of an $ab$-path in $E(G) \setminus E(B)$ plus a $bp$-path in $E(B)$ for some node $p$.

Consider now an external path $P$ of $(B, F, a, b)$ and a different patch $(\tilde B, \tilde F, \tilde a, \tilde b)$.
Observe that $P$ either avoids $E(\tilde B)$ entirely, or it has an $\tilde a \tilde b$-subpath in $E(\tilde B)$.
We say paths of the latter type {\em transit} $\tilde B$.
We let $x_C(\tilde B)$ be the sum of flow on (external) support paths of $x_C$ that transit $\tilde B$.

It will be useful for us to impose some additional structure on our stellar-planar instance.
We say a feasible stellar-planar instance is {\em friendly} if there exists a feasible flow $x$ such that, for every patch $C = (B, F, a, b)$, one of the following holds:
\begin{enumerate}
    \item $E(B) = \emptyset$ (i.e. $C$ consists only of demand between $a$ and $b$),
    \item $F = \emptyset$ (i.e. $C$ consists only of supply edges), or
    \item $B$ is connected and $x_{C'}(B) = 0$ for all other patches $C'$ (i.e. no other patch routes flow through $C$).
\end{enumerate}



    

We say a pinning $(H', d')$ is {\em good} if every demand edge is either parallel to a supply edge in $G$ or is between the two attachment nodes of some patch.
In particular, if a pinning is good then $G + H'$ is planar.

\begin{theorem}
    \label{thm:unsplittable-pinning-friendly}
    Let $(G, H, u, d)$ be a feasible stellar-planar instance.
    Additionally, suppose the instance is friendly with respect to a feasible flow $x$.
    There exists a good pinning $(H', d')$ of the demands such that $(G, H', u^*, d')$ is feasible, where $u^*(e) \le u(e) + 2 d_{\max}$.
\end{theorem}
\begin{proof}
    Consider a counterexample $\mathcal I = (G, H, u, d)$ that minimizes the number of patches.

    If every patch $C = (B, F, a, b)$ has either $E(B) = \emptyset$ or $F = \emptyset$, then the instance is fully-planar and we are already done.
    So, since the instance is friendly, there exists a patch $C = (B, F, a, b)$ such that $B$ is connected and $x_{C'}(B) = 0$ for all other patches $C'$ (i.e. no other patch routes flow through $C$).


    Let $\gamma$ be the sum of flow that $x_C$ routes on external paths (possibly $\gamma = 0$).
    Define the following two multiflow instances.
    \begin{itemize}
        \item Let $\mathcal J(\overline C) = (\overline B, \overline F, u^{(1)}_{\overline C}, d_{\overline C})$, where $u^{(1)}_{\overline C}(e) = x_{\overline C}(e)$ for all $e \in E(\overline B)$ and $d_{\overline C}$ is $d$ restricted to $\overline F$.
        Observe that this instance is also feasible and friendly.
        Since it has fewer patches than $\mathcal I$, by minimality there exists a good pinning $(\overline {F'}, d'_{\overline C})$ such that $(\overline B, \overline {F'}, u^{(2)}_{\overline C}, d'_{\overline C})$ is feasible where $u^{(2)}_{\overline C}(e) = x_{\overline C}(e) + 2 d_{\max}$ for all $e \in E(\overline B)$.
        \item Let $\mathcal J(C) = (B', F, u^{(1)}_C, d_C)$ where $B'$ is $B$ plus an extra supply edge $e^* = ab$, $u^{(1)}_C(e) = x_C(e)$ for all $e \in E(B)$, $u^{(1)}_C(e^*) = \gamma$, and $d_C$ is $d$ restricted to $F$.
        Observe that this instance is feasible.
        Since $\mathcal J(C)$ is single-source, it has an unsplittable flow with additive error $d_{\max}$.
        By pinning along these unsplittable flow paths, we find a pinning $(F', d'_C)$ such that each demand is pinned only along supply edges of $B$ or between $a$ and $b$, and $(B', F', u^{(2)}_C, d'_C)$ is feasible where $u^{(2)}_C(e) = x_C(e) + d_{\max}$ for $e \in E(B)$ and $u^{(2)}_C(e^*) = \gamma + d_{\max}$.
    \end{itemize}
    
    
    Let $P$ be an $ab$-path in $B$, which exists because $B$ is connected.
    Now, let $u^{(3)}_C$ be obtained from $u^{(2)}_C$ by decreasing the capacity of $e^*$ by $d_{\max}$ (to $\gamma$) and increasing the capacity of all edges of $P$ by $d_{\max}$ (to $x_C(e) + 2d_{\max}$).
    Observe that $(B', F', u^{(3)}_C, d'_C)$ is still feasible.
    Now, we remove the added edge $e^*$ and restore the edges of $\overline B$ to the instance.
    Consider $(G, F', u^{(4)}_C, d'_C)$ where $u^{(4)}_C(e) = u^{(3)}_C(e) \le x_C(e) + 2d_{\max}$ for $e \in E(B)$ and $u^{(4)}_C(e) = x_C(e)$ for $e \in E(\overline B)$.
    This instance is again feasible, since $x_C$ supports routing $\gamma$ units of flow from $a$ to $b$ in $\overline B$.
    
    Consider then $(F' \cup \overline{F'}, d'_C + d'_{\overline C})$, which is a pinning of the original $(H, d)$.
    We claim this is a pinning with the desired properties.
    For capacities $u^* = u^{(4)}_C + u^{(2)}_{\overline C}$, we have that the instance $(G, F' \cup \overline{F'}, u^*, d'_C + d'_{\overline C})$ is feasible.
    For $e \in E(B)$ we have $u^*(e) = u^{(4)}_C(e) \le x_C(e) + 2d_{\max} \le u(e) + 2d_{\max}$.
    For $e \in E({\overline B})$ we have that $u^*(e) = u^{(4)}_{C}(e) + u^{(2)}_{\overline C}(e) = x_C(e) + x_{\overline C}(e) + 2d_{\max} \le u(e) + 2d_{\max}$. 
    Additionally, the pinning is good because $(\overline {F'}, d'_{\overline C})$ is good and $(F', d'_C)$ pins only along supply edges of $B$ or between $a$ and $b$.
    Hence this pinning has the desired properties.
\end{proof}

Our proof above is most reminiscent of the proof in \cite{aleman2025unsplittable}.
They similarly select a subset of demands to pin at each step and consider the sub-instance with just those demands and their used capacity in the whole graph.
However, their steps are considerably more complicated.
The selected demands are those internal to a given {\em ear} of the outerplanar supply graph.
To pin those demands, they show how to compress the sub-instance into a ring instance that is still feasible, then apply results that give low-congestion unsplittable flow in rings \cite{skutella2016note,daubel_improved_2022} to find a pinning.
This compression step is not trivial and requires carefully assigning capacity by looking at the dual of the outerplanar graph.
For stellar-planar instances, we can use that each patch is single-source to find a pinning of its demands.
Once this local pinning is done, \cite{aleman2025unsplittable} shows how to transfer error accumulated outside of the ear back onto the ear so that the accumulated error is not too large in future iterations, much as we do here by transferring error accumulated in $\overline B$ back to $B$.

Of course, we are not finished, as we need to show how to reduce to a friendly instance.
However, it turns out this error transferring idea ends up being useful again.


\begin{theorem}
    \label{thm:stellar-planar-pinning}
    Let $(G, H, u, d)$ be a feasible stellar-planar instance.
    There exists a pinning $(H', d')$ such that $G + H'$ is planar and $(G, H', u^*, d')$ is feasible where $u^*(e) = u(e) + 4 d_{\max}$.
\end{theorem}
\begin{proof}
    First, we show that if a patch $C = (B, F, a, b)$ has any supply edges ($E(B) \neq \emptyset$), then we may assume that $B$ is connected.
    Suppose $B$ is not connected, and in particular $a$ and $b$ lie in different components $B_a$ and $B_b$.
    We may assume $B$ contains no other components, since such components would not be connected to the rest of the graph and have no incident demands, so they can be deleted safely.
    If $B_b$ does not have any endpoints of demands of $F$, then we can delete the edges of $B_b$ (which can be done without affecting feasibility) and replace $C$ with $C' = (B_a, F, a, b')$ where $b'$ is an arbitrary other node on the outside face of $B_a$.
    So suppose there are demands with endpoints in $B_b$.
    For each such demand $ap$ where $p \neq b$, we pin the demand with $ab, bp$.
    This pinning does not affect feasibility because every $ap$-path must use $b$.
    The resulting instance is still stellar-planar, but with a slightly different decomposition into patches: $C$ instead is split into three patches $C_a, C_b, C_0$.
    The patch $C_a$ has supply graph $B_a$, the demands of $F$ whose endpoints all lie in $B_a$, source $a$, and an arbitrary other node on the outside face of $B_a$ as its other attachment node.
    The patch $C_b$ has supply graph $B_b$, the new demands from $b$ to nodes in $B_b$, source node $b$, and an arbitrary other node on the outside face of $B_b$ as its other attachment node.
    The patch $C_0$ has a supply graph with no edges, a demand graph only with edges from $a$ to $b$, source $a$, and other attachment node $b$.
    
    We now construct a friendly instance as follows. Let $x$ be any feasible flow.
    Consider a patch $C = (B, F, a, b)$ that has at least one demand edge and at least one supply edge.
    We have that $B$ is connected.
    If $C$ satisfies that $x_{C'}(B) = 0$ for all other patches $C'$, we do nothing with $C$.
    Otherwise, we call $C$ a {\em bad patch}.
    Let $\gamma = \sum_{C' \neq C} x_{C'}(B) > 0$.
    We add a new supply edge $ab$ with capacity equal to $\gamma$.
    Additionally, we decrease the capacity of every edge $e$ of $B$ by $\sum_{C' \neq C} x_{C'}(e)$.
    We put the new supply edge $ab$ into its own new patch $C^+$ that consists of solely this supply edge with no demand edges.
    Let $(G', H, u^{(1)}, d)$ be the resulting instance after performing this operation for all bad patches $C$.
    This instance is still feasible and stellar-planar, and it is also friendly: we can modify $x$ to a feasible flow for the new instance by redirecting all flow that transits $C$ from other patches and sending it through $C^+$ instead.

    By \Cref{thm:unsplittable-pinning-friendly}, there exists a pinning $(H', d')$ such that $G' + H'$ is planar and $(G', H', u^{(2)}, d')$ is feasible where $u^{(2)}(e) = u^{(1)}(e) + 2 d_{\max}$.
    We show how to cast this back to $G$.
    Intuitively, we transfer error from the new edges back to the bad patches they originated from.
    We still have that $G + H'$ is planar.
    Let $u^*$ be capacities on $G$ defined as follows.
    For edge $e \in E(G)$, if $e$ is not part of a bad patch, let $u^*(e) = u^{(2)}(e) = u^{(1)}(e) + 2d_{\max} = u(e) + 2d_{\max}$.
    If $e$ is part of a bad patch $C = (B, F, a, b)$, let $u^*(e) = u^{(2)}(e) + \sum_{C' \neq C} x_{C'}(e) +2d_{\max} = u^{(1)}(e) + \sum_{C' \neq C} x_{C'}(e) + 4 d_{\max} = u(e) + 4d_{\max}$.
    This instance is still feasible, because we can take a feasible flow for $(G', H', u^{(2)}, d')$ and transfer the flow routed on the newly introduced edges back through their originating bad patch.
    Hence this pinning satisfies the desired properties.
\end{proof}

Given \Cref{thm:stellar-planar-pinning}, \Cref{thm:unsplittable} follows easily by using the $2d_{\max}$ error unsplittable flow result of \cite{espinosa2026unsplittable} for fully-planar instances.

\section{Future Work}

We believe we can show that stellar-planar instances also have a property called {\em uncrossability}.
An instance is uncrossable if any feasible flow can be converted into one such that the curves identified by its support paths (paths with $x(P) > 0$) do not cross each other in the plane.
Planar single-source and fully-planar instances are known to be uncrossable, and for the latter this has been used to design constant-factor approximation algorithms for maximum edge-disjoint paths (MEDP) \cite{huang2021approximation,garg2022integer,schlomberg_packing_2025}.
Recent work \cite{chekuri_uncrossed_2026} has shown that any uncrossable instance family admits a constant-factor approximation algorithm for MEDP (provided the uncrossing can be made efficient), and so stellar-planar instances would inherit such an algorithm.

\section*{Acknowledgments}
We thank an anonymous SOSA reviewer for suggestions to improve the induction step in \Cref{thm:unsplittable-pinning-friendly}.
Claude Opus 5.5 was used to assist in creating \Cref{fig:stellar-planar-concrete} and minor proofreading. Generative AI was not used in any other substantive way in this work.
The authors assume responsibility for all content.







\bibliography{mybibliography}

\end{document}